\documentclass[aps,prb,preprint,groupedaddress]{revtex4}
\usepackage{graphicx}
\usepackage{textcomp}
\usepackage{gensymb}
\usepackage{amsmath}

\begin{document}


\title{Molybdenum-Rhenium alloy based high-$Q$ superconducting microwave resonators}


\author{Vibhor~Singh}
\email{v.singh@tudelft.nl}
\author{Ben~H.~Schneider}
\author{Sal~J.~Bosman}
\author{Evert~P.~J.~Merkx}
\author{Gary~A.~Steele}


\affiliation{Kavli Institute of NanoScience, Delft University of Technology, PO Box 5046, 2600 GA, Delft, The Netherlands.}


\date{\today}

\begin{abstract}

Superconducting microwave resonators (SMR) with high quality factors have become an important technology in a wide range of applications. 
Molybdenum-Rhenium (MoRe) is a disordered superconducting alloy with a noble surface chemistry and a relatively high transition temperature. These properties make it attractive for SMR applications, but characterization of MoRe SMR has not yet been reported. 
Here, we present the fabrication and characterization of SMR fabricated with a MoRe 60-40 alloy.
At low drive powers, we observe internal quality-factors as high as 700,000. Temperature and power dependence of the internal quality-factors suggest the presence of the two level systems from the dielectric substrate dominating the internal loss at low temperatures.
We further test the compatibility of these resonators with high temperature processes such as for carbon nanotube CVD growth, and their performance in the magnetic field, an important characterization for hybrid systems. 
\end{abstract}


\maketitle



Molybdenum-Rhenium (MoRe)  is an attractive superconducting alloy  which has been explored as early as the 1960s \cite{lerner_magnetic_1967,yasaitis_microwave_1975,andreone_mo-re_1989}.
Recently, there has been a renewed interest in MoRe superconductors due to its electrical and mechanical properties \cite{leonhardt_investigation_1999,seleznev_deposition_2008,sundar_electrical_2013,aziz_molybdenum-rhenium_2014}.
Depending on the alloying ratio and film deposition temperatures, thin films of MoRe have shown superconducting transition temperatures ($T_c$) ranging from 8~K to 13~K and residual resistance ratios $RRR\approx 1$  indicating their highly disordered nature.
Despite these attractive properties, it has not been explored as a material for the superconducting coplanar waveguide resonators.
The High frequency and low dissipation of superconducting microwave resonators (SMR) have led to their use in a variety of applications ranging from sensitive photon detectors \cite{day_broadband_2003}, quantum computation \cite{wallraff_strong_2004,palacios-laloy_tunable_2008}, and coupling to nanoelectromechanical resonators \cite{regal_measuring_2008}.
There has been also a considerable interest in combining them to DC electron transport devices based on carbon nanotubes \cite{delbecq_photon-mediated_2013}, nanowires \cite{petersson_circuit_2012}, 2-dimensional electron gases \cite{frey_dipole_2012,toida_vacuum_2013}, spin ensembles \cite{kubo_strong_2010,amsuss_cavity_2011}, and magnons \cite{huebl_high_2013}.
These hybrid coupling devices require robustness of low dissipation in the SMR against the fabrication process (such as high-temperature growth) and the measurement scheme, such as in magnetic field. 
The versatile use of SMR and constraints for various applications have led to the studies of different superconductors \cite{barends_niobium_2007,hammer_superconducting_2007,barends_contribution_2008}, where MoRe could be an attractive alternative.

The motivation for exploring MoRe for the SMR is multifold.
Its highly disordered nature of MoRe makes it attractive for kinetic inductance detectors \cite{day_broadband_2003}.
It makes highly transparent superconducting contacts with carbon nanotubes \cite{schneider_coupling_2012} and has been used to make contact with graphene in forming a high quality-factor superconducting opto-mechanical device \cite{singh_optomechanical_2014}.
Furthermore, a high upper critical magnetic field \cite{sundar_electrical_2013} makes it attractive for applications requiring magnetic field. 
Here, we investigate SMR fabricated using a 60-40 alloy of MoRe.
We provide a fabrication recipe to make superconducting microwave resonators with low dissipation, characterize their performance after a high temperature chemical vapor deposition process and in the presence of magnetic field.

\begin{figure}
\includegraphics[width=120mm]{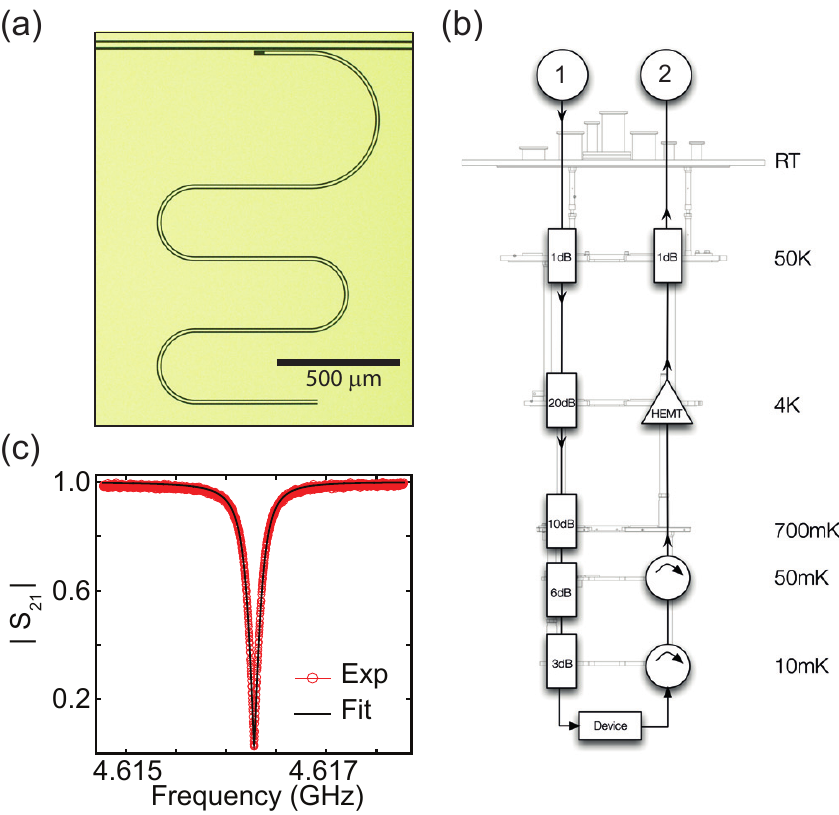}
\caption{\textbf{Basic device characterization:} (a) False color optical microscope image of a quarter-wavelength resonator side coupled to a transmission line in the coplanar waveguide geometry. The bright region represents the deposited metal and the dark region is the dielectric substrate (sapphire). (b) Schematic of the setup for low temperature measurements. Microwave signal reaching to the device is sufficiently attenuated to equilibrate at various temperature stages. (c) Experimentally measured transmission coefficient $|S_{21}|$ for a resonator at $T=$~830~mK (red circles) with $\approx$100 probe photons. The black curve is the fit to data to extract the resonator parameters yielding $Q_i\sim 700,000$.}\label{fig1}
\end{figure}

The SMR were designed in a coplanar waveguide geometry and fabricated on a sapphire wafer (substrate thickness $\sim$ 430~$\mu$m) in order to minimize dielectric losses.
As cleaning of the substrate surface seems to play an important role in minimizing two-level systems \cite{megrant_planar_2012}, an extensive cleaning of the sapphire wafer
is performed with phosphoric acid (H$_3$PO$_4$) at $75~\celsius$ for 30 minutes followed by rinsing in DI water for 2 hours.
After exposing the fresh surface of the wafer, it was immediately loaded in the vacuum chamber for MoRe film deposition.
Using an RF sputtering system, we deposit a 145~nm thick MoRe film with a continuous flow of Ar (pressure $1.5\times 10^{-3}$~mTorr) from a $\sim99.95~\%$ purity, single target of MoRe.
The SMR designs were patterned using e-beam lithography on a three layer mask (S1813/W(Tungsten)/PMMA-950) followed by the etching of MoRe by SF$_6$/He plasma.
We use a frequency multiplexing scheme to side-couple multiple quarter wavelength resonators of different frequencies to a common transmission line.
Figure~\ref{fig1}(a) shows an optical microscope image of such a resonator after the fabrication process. The quarter-wavelength coplanar waveguide resonator is formed by terminating a transmission line (characteristic impedance of 50~$\Omega$ and 10~$\mu$m wide trace) to the ground plane. For microwave measurements, the samples were mounted in a light-tight microwave box and were cooled down in a dilution fridge or a He-3 cryostat with sufficient attenuation at each temperature stage to thermalize the microwave photons reaching the sample.
A schematic of the attenuation scheme in the dilution refrigerator is shown in Figure~\ref{fig1}(b). For these sputtered thin films, 
we measure a typical room temperature resistivity of 88~$\mu \Omega$-cm, $RRR\sim1.2$ and $T_c\sim 9.2$~K.

The low power transmission response $S_{21}$ for a side-coupled quarter wavelength resonator near its resonance frequency $f_0$ can be modeled by
\begin{align}
S_{21} (f)=\frac{S_{21}^{min}+2iQ_l\delta f}{1+2iQ_l\delta f},
\end{align}
where $S_{21}^{min} = Q_c/(Q_c+Q_i)$ is the swing in transmission at resonance, $\delta f = \frac{f-f_0}{f_0}$, $Q_l =Q_c\times Q_i/(Q_c+Q_i)$ is the loaded quality-factor, $Q_c$ is the coupling quality-factor and $Q_i$ is the internal quality-factor of the resonator.
Figure~\ref{fig1}(c) shows the measurement of the transmission coefficient $S_{21}$ for a resonator along with the fitted curve at $T=830$~mK.
We observe an internal quality-factor of 700,000 at low probe powers, comparable to the best results using a highly optimized Al deposition \cite{megrant_planar_2012}.

\begin{figure}
\includegraphics[width=90mm]{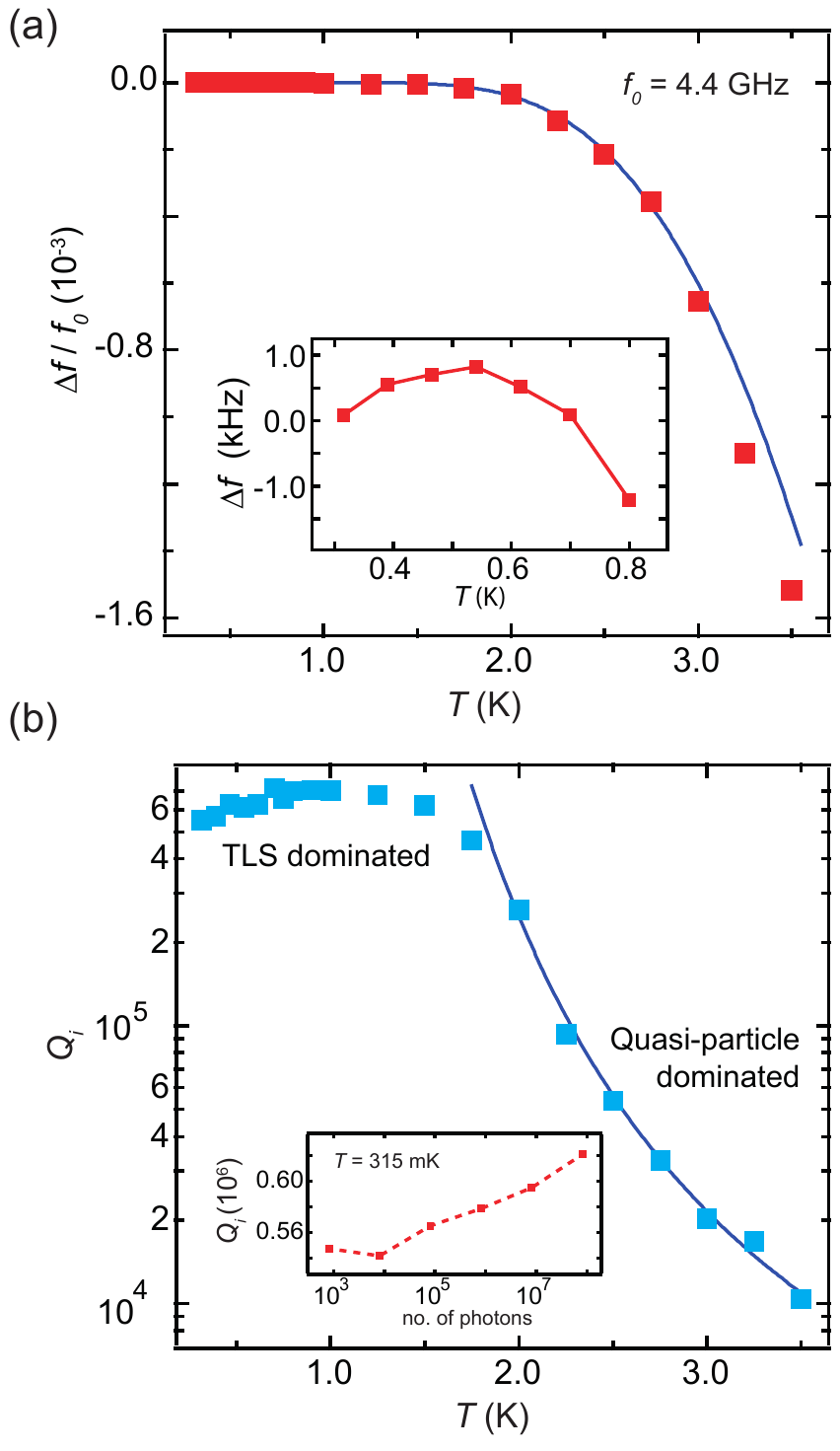}
\caption{\textbf{Characterization with temperature and identifying the effects of two level systems:}  (a) Normalized shift in the resonance frequency of different resonators with temperature. The blue line is the expected behavior from Mattis-Bardeen (M-B) theory (no fit). The inset shows the zoomed-in view of typical resonance frequency shift in the low temperature range. (b) Temperature dependence of the internal quality-factor $Q_i$ at a drive power corresponding to $\approx1000$ photons. The blue line represents the losses due to quasi-particles calculated using M-B theory without any fit parameters. Inset shows the dependence of $Q_i$ with the number of probe photons at $T=315$~mK suggesting the presence of two level systems.  \label{fig2} }
\end{figure}

Figure~\ref{fig2} shows the typical temperature dependence of the normalized resonance frequency shift $\Delta f /f_0 =  \frac{f_0(T)-f_0\text{(315~mK)}}{f_0{\text{(315~mK)}}}$ and the internal quality factor $Q_i$ for a resonator.
With an increase in temperature above $\sim$2~K, the reduced cooper-pair density in the superconductor presents a larger kinetic inductance resulting in a sharp drop of the resonator frequency as temperature approaches $T_c$ as shown in Figure~\ref{fig2}(a). The corresponding increase in the number of quasi-particles also
leads to a sharp drop in the internal quality-factor of the resonator as shown in Figure~\ref{fig2}(b). The frequency and internal quality-factor response with temperature in the vicinity of $T_c$ can be very well described by the Mattis-Bardeen (M-B) theory \cite{mattis_theory_1958}, which predicts $\Delta f /f_0 = \frac{\alpha}{2}\frac{\Delta\sigma_2}{\sigma_2}$ and $Q_i = \frac{2}{\alpha}\frac{\sigma_2}{\sigma_1}$, where $\alpha$ is the kindetic inductance fraction, $\sigma_1$ and $\sigma_2$ are the real and imaginary parts of the complex conductivity, respectively. 
Using the experimentally measured normal state resistivity and $T_c$, we estimated a surface inductance $L_s\sim$0.35~pH for these films.
The blue curve in Figure~\ref{fig2}(a) and (b) shows the expected temperature dependence of the normalized resonance frequency shift and internal quality-factor calculated using the M-B theory with no free parameters.
For temperatures below 2~K, the resonance frequency shift and saturation in the internal quality-factor are not captured by the M-B theory as it predicts no change in the resonance frequency for $T\ll T_c$ and exponentially large $Q_i$.
In this temperature range, we measure a small increase in the resonance frequency as shown in the inset of Figure~\ref{fig2}(a).
This small positive shift is attributed to the presence of two-level systems (TLS) and can be written as
$\frac{\Delta f}{f_0} = \frac{F}{2}\frac{\Delta \epsilon}{\epsilon}$,
where $F$ is a geometrical factor and $\epsilon$ is the dielectric constant dominated by the resonant interactions over the relaxation of two-level systems at low temperature \cite{phillips_two-level_1987,gao_experimental_2008}. 
Correspondingly, $Q_i$ saturates and follows a trend similar to the normalized resonance frequency shift indicating the  presence of TLS. With a large number of  probe photons, the damping from the TLS saturates and leads to an increase in the internal quality-factor as shown in the inset of Figure~\ref{fig2}(b) (additional data provided in supplemental material (SM) \cite{SM}).

\begin{figure}
\includegraphics[width=90mm]{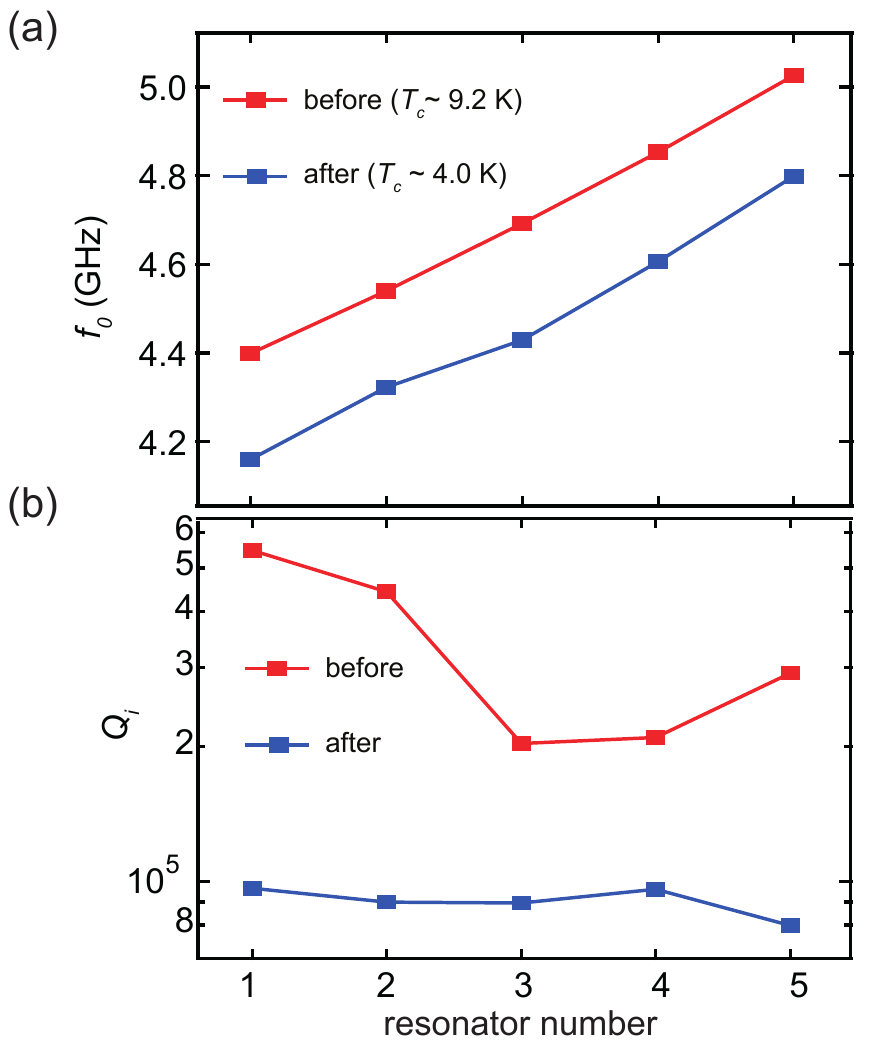}
\caption{\textbf{Effect of high temperature CVD process on MoRe resonators measured at 315~mK:} Comparison of (a) resonance frequency and (b) internal quality-factors of 5 resonators after first annealing them at $900\,^{\circ}{\rm C}$ in hydrogen flow and then flowing methane for 10 minutes. After annealing superconducting transition temperature decreases from 9.2~K to 4.0~K and room temperature resistivity changes from 88~$\mu \Omega$-cm to 108~$\mu \Omega$-cm. \label{fig3}}
\end{figure}

After establishing the large quality-factors and role of the two level systems in these resonators, we explore the changes in characteristics of the resonators after a chemical vapor deposition process at high temperatures. In order to form hybrid systems with the SMR by coupling them to other materials like a carbon nanotube based quantum dots and mechanical resonators, often a high temperature growth process is required. 
To explore the compatibility of these microwave resonators with high-temperature growth processes, we subject them to a chemical vapor deposition (CVD) growth process used to grow carbon-nanotubes. A continuous hydrogen flow os present during this process. Once the oven temperature of $900\,^{\circ}{\rm C}$ is reached, a methane (CH$_4$) flow was added for 10 minutes (see SM for details \cite{SM}). After this process, the superconducting transition temperature of these films decreased to $T_c \sim 4.0$ K with a room temperature resistivity of 108~$\mu\Omega$-cm.
Figure~\ref{fig3} shows a typical change in resonator characteristics before and after high temperature annealing measured at 315~mK.
For all resonators, we have observed an approximately 250 MHz decrease in the resonance frequency after annealing as shown in Figure~\ref{fig3}(a).
Such a decrease in the resonance frequency can be attributed to the increased kinetic inductance fraction due to the degradation in the superconducting film quality.
After this process, we also observe a reduction in the internal quality-factor of the resonator, shown in Figure~\ref{fig3}(b). The larger drop of $Q_i$ after annealing suggests increased dielectric losses in the sapphire substrate (additional data provided in supplemental material \cite{SM}). We have also studied the effects of annealing the resonators in pure hydrogen at $900\,^{\circ}{\rm C}$, and find that the $T_c$ of the film increases to 11.8~K (improving the quality of the superconducting film). However, annealing also lead to increased dielectric loss in the substrate (see SM for more details).

\begin{figure}
\includegraphics[width=90mm]{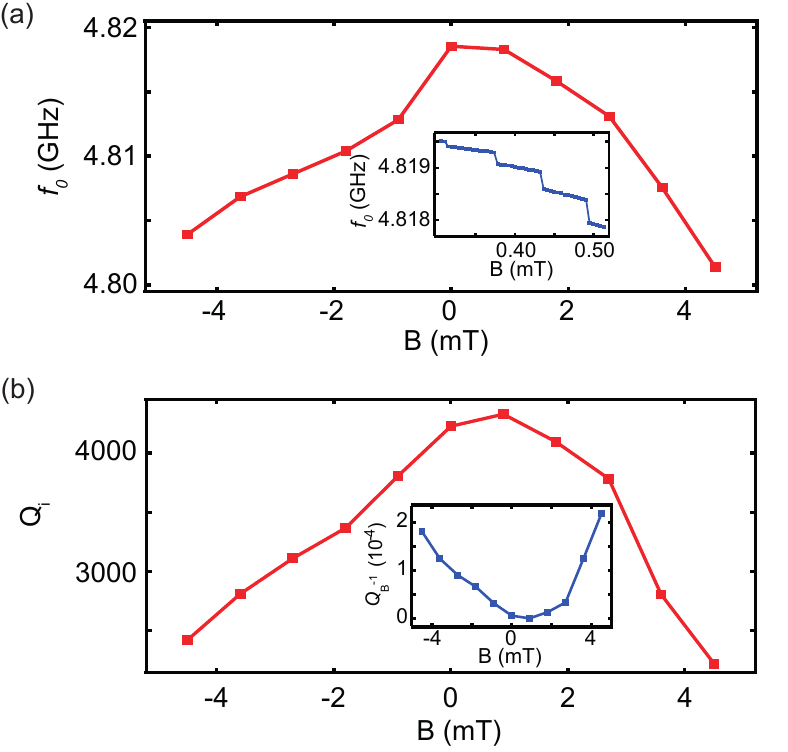}
\caption{\textbf{Characterization in the magnetic field at 4.2~K:} (a) Plot of resonance frequency with magnetic field at 4.2~K showing a drop in resonance frequency. Inset shows the detailed measurement of resonance frequency ($f_0$) in a narrow range of magnetic field to highlight the discrete jumps observed in resonance frequency due to the incorporation of vortices. (b) Plot of internal quality-factor $Q_i$ with magnetic field at 4.2~K. The internal quality-factor drops in the presence of magnetic field. Inset shows the magnetic field contribution to the total dissipation in the resonator. \label{fig4}}
\end{figure}

For various hybrid coupling schemes of microwave resonator to systems such as spin ensembles \cite{kubo_strong_2010,amsuss_cavity_2011}, magnons \cite{huebl_high_2013}, and nanotube SQUIDs \cite{schneider_coupling_2012}, a low dissipation of the SMR in a magnetic field is desirable. 
In a magnetic field, formation of flux vortices can lead to added dissipation \cite{song_microwave_2009,bothner_magnetic_2012}.
After characterizing the high temperature annealed resonators, we subject them to small magnetic fields up to $\sim$~4.5~mT.
Figure~\ref{fig4}(a) shows the resonance frequency of one of the resonator with magnetic field applied perpendicular to the sample surface at 4.2~K. The presence of the vortices adds a reactive part to the resonator's  microwave response and hence reduces its resonance frequency. This becomes quite evident by looking at the detailed measurement of the resonance frequency with small steps in the magnetic field. The incorporation of individual vortices is reflected as abrupt jumps in the resonance frequency as shown in the inset.

The formation of vortices not only adds a reactive part to the microwave response, but also adds a dissipative part to it. Figure~\ref{fig4}(b) shows the plot of reduction in $Q_i$ with magnetic field (corresponding to the same resonator as shown in Figure~\ref{fig4}(a)). Application of the magnetic field opens up an extra dissipation channel due to the motion of vortices in response to the microwave field, and it is often characterized by a magnetic field dependent dissipation rate given by $Q_B^{-1}=Q_i^{-1}(B)-Q_i^{-1}(0)$ \cite{song_microwave_2009,bothner_magnetic_2012}. 
The inset of figure~\ref{fig4}(b) shows the plot of the magnetic field contribution to the total dissipation rate, which is smaller than in the previously studied devices with similar geometry using rhenium \cite{song_microwave_2009}.
It should be noted here that the geometry of the SMR studied here has continuous ground plane metalization, which leads to a significant magnetic flux focusing between the central conductor and the ground plane. The susceptibility to the magnetic field can further be improved by making holes in the ground plane, which act as vortex pinning centers \cite{song_reducing_2009,bothner_improving_2011,graaf_magnetic_2012}.

In summary, we presented a fabrication method of superconducting microwave resonators based on MoRe alloy and observe high internal quality-factors.
We characterized its surface impedance, role of the two-level systems and  compatibility of these resonators with high temperature CVD processes.
Post-CVD high internal quality-factor ($\approx 10^5$) and compatibility in the magnetic would lead to flux tunable hybrid systems by directly incorporating these resonators in growth processes with carbon nanotubes and nanowires, leading to new research directions \cite{schneider_coupling_2012,souquet_photon-assisted_2014}.

\section*{acknowledgments}
The authors would like to thank Prof. Teun Klapwijk, Akira Endo and Pieter de Visser for discussions and help during the measurements. The work was supported by the Dutch Organization for Fundamental Research on Matter (FOM).


\newpage

\end{document}